\documentclass[
    reprint,
    amsmath,
    amssymb,
    aps,
    prmaterials,
    superscriptaddress
]{revtex4-2}

\usepackage{graphicx}
\usepackage{dcolumn}
\usepackage{bm}
\usepackage{color}
\usepackage{soul}
\usepackage{makecell}
\usepackage{mathrsfs}
\usepackage{booktabs}
\usepackage{afterpage}
\usepackage{pifont}
\usepackage[version=3]{mhchem}
\usepackage[caption=false]{subfig}
\usepackage{xtab}
\usepackage{array}
\usepackage{multirow}
\usepackage[colorlinks=true, allcolors=blue]{hyperref}

\begin{document}

\title{Decoupling Exciton Energetics from Wavefunction Geometry in Imide-Based Defect Double Antiperovskites}

\author{Dhriti Mandal}
\email{dhriti@physics.iitd.ac.in}

\author{Saswata Bhattacharya}
\email{saswata@physics.iitd.ac.in}

\affiliation{Department of Physics, Indian Institute of Technology Delhi,
New Delhi 110016, India}

\begin{abstract}
	The excited-state properties of imide-based defect double antiperovskites AE$_5$AsPn(NH)$_2$ (AE = Ca, Sr; Pn = Sb, Bi) are investigated using spin--orbit-coupled quasiparticle calculations, the model Bethe--Salpeter equation (mBSE), and analytical descriptions of lattice-polarization effects. Across five tetragonal ($P4/mmm$) and monoclinic ($Cm$) structural models, the calculated exciton binding energies range from 140 to 210 meV. Chemical substitution modifies both the binding energies and the hydrogenic estimates of exciton spatial extent. By contrast, the structural distortion from tetragonal $P4/mmm$ ($D_{4h}$) to monoclinic $Cm$ ($C_s$), which removes inversion and fourfold rotational symmetry, produces comparatively modest changes in binding energy but appreciably reshapes the real- and momentum-space exciton distributions. This contrast reveals a partial decoupling between exciton energetics and wavefunction geometry, and identifies a symmetry-associated sensitivity of exciton geometry to structural distortion. The changes are understood to occur together with modifications of local coordination and electronic dispersion, rather than as an isolated effect of symmetry alone. Lattice polarization provides an additional screening channel, reducing the estimated exciton binding energies by approximately 9--16\%. Complementary Fr\"ohlich-polaron calculations yield moderate mass enhancement and model-estimated electron-polaron mobilities of 39--99 cm$^2$ V$^{-1}$ s$^{-1}$, with comparatively modest differences between the two polymorphs. These results highlight the distinct responses of exciton energetics, exciton wavefunction geometry, and carrier transport to chemical and structural modifications.
\end{abstract}

\maketitle

\section{Introduction}

Metal-halide perovskites have attracted extensive research interest because of their strong optical absorption, tunable band gaps, and favorable carrier-transport characteristics~\cite{kojima2009organometal,stranks2015metal,green2014ho,snaith2018present}. Their technological applications, however, are constrained by concerns including chemical and thermal instability and, for many leading compositions, the presence of toxic lead~\cite{babayigit2016toxicity,berhe2016organometal,ju2018earth}. These limitations have motivated the search for alternative semiconductors with different anion chemistries and robust inorganic frameworks~\cite{kobayashi2018property,krivovichev2024structural,grosso2023accessible}. Nitride-based perovskites and antiperovskites are of particular interest because their chemically distinct anionic networks provide additional opportunities to tune electronic structures through cation substitution and structural modification~\cite{monga2024theoretical,zhong2021structure,grosso2023accessible}.

Antiperovskites are characterized by an arrangement of cations and anions that is inverted relative to the conventional perovskite structure~\cite{krivovichev2024structural,zheng2021antiperovskite}. In the recently reported defect-imide double antiperovskites AE$_5$AsPn(NH)$_2$ (AE = Ca, Sr; Pn = Sb, Bi), the incorporation of imide groups is accompanied by ordered alkaline-earth vacancies~\cite{chau2025defect}. The resulting structures contain distinct As- and Pn-centered coordination environments together with imide-containing alkaline-earth units~\cite{chau2025defect}. The experimentally reported compounds exhibit direct optical gaps in the visible range, with electronic states from several crystallographic sublattices contributing to the band edges~\cite{chau2025defect}. The combination of vacancy ordering, chemically distinct anionic units, and structural polymorphism therefore makes this family an interesting platform for investigating how composition and crystal structure influence electronic and excited-state properties~\cite{monga2024theoretical,zhong2021structure,chau2025defect}.

Recent experimental and theoretical studies established the synthesis, crystal structures, and baseline optoelectronic properties of Ca$_5$AsSb(NH)$_2$, Ca$_5$AsBi(NH)$_2$, and Sr$_5$AsBi(NH)$_2$, including tetragonal ($P4/mmm$) and monoclinic ($Cm$) polymorphs for selected compositions~\cite{chau2025defect}. While hybrid-functional calculations, hydrogenic Wannier--Mott exciton estimates (28--48~meV), and carrier-mobility estimates in Ref.~\cite{chau2025defect} identified these compounds as promising solar absorbers, a more complete description of their excited-state properties requires going beyond single-particle electronic structures and hydrogenic effective-mass estimates. The presence of heavy elements such as Sb and Bi makes spin--orbit coupling (SOC) essential for describing the band-edge states~\cite{kurita2020systematic}, whereas quasiparticle self-energy corrections substantially modify band gaps and orbital ordering beyond conventional density-functional theory~\cite{onida2002electronic,hybertsen1986electron}. Furthermore, explicit electron--hole interactions govern both the energies and wavefunctions of optical excitations, requiring a clear distinction between the fundamental quasiparticle gap and the optical excitation energy~\cite{onida2002electronic,rohlfing1998electron,rohlfing2000electron,bechstedt2016many}.

An additional consideration in these materials is the influence of lattice polarization. In polar semiconductors, the ionic contribution to dielectric screening modifies the effective electron--hole interaction and can reduce exciton binding energies~\cite{filip2021phonon,sak1972effective}. The same polar lattice also couples to free charge carriers through long-range Fr\"ohlich interactions, leading to polaron formation and renormalization of carrier effective masses, with consequent implications for transport~\cite{feynman1955slow,feynman1962mobility,devreese2009frohlich}. These two manifestations of lattice polarization concern different aspects of the excited-state response: ionic screening modifies the interaction between an electron and a hole, whereas carrier--phonon coupling dresses individual charge carriers. Examining both effects within the same material family can therefore provide a more complete understanding of the role of lattice dynamics in excitonic and carrier properties.

Beyond chemical composition, structural polymorphism provides a route to modifying the symmetry constraints and local electronic environment experienced by an exciton. The tetragonal $P4/mmm$ structure has point group $4/mmm$ ($D_{4h}$), including inversion and fourfold rotational symmetry, whereas the monoclinic $Cm$ structure has point group $m$ ($C_s$), lacking both inversion and the fourfold axis~\cite{chau2025defect}. This symmetry reduction changes the symmetry classification of electronic states and the constraints on their mixing; together with the accompanying changes in local coordination and band dispersion, it can alter the directional character of electron--hole correlations. Importantly, the removal of inversion symmetry does not by itself establish enhanced optical activity: the transition matrix elements must be evaluated explicitly. A comparison of polymorphs with the same composition therefore provides a useful way to examine how a structural distortion associated with symmetry lowering affects exciton energies and wavefunction distributions. Despite the experimental identification of these polymorphs~\cite{chau2025defect}, their comparative quasiparticle, excitonic, and polaronic properties have not been systematically investigated.

In this work, we present a first-principles investigation of the tetragonal $P4/mmm$ structures of Ca$_5$AsSb(NH)$_2$, Ca$_5$AsBi(NH)$_2$, and Sr$_5$AsBi(NH)$_2$, together with the monoclinic $Cm$ structures of Ca$_5$AsSb(NH)$_2$ and Sr$_5$AsBi(NH)$_2$. After examining their structural characteristics, formation energetics, elastic response, and lattice dynamics, we investigate the electronic structures using SOC-inclusive quasiparticle calculations within the $G_0W_0$ approximation~\cite{hybertsen1986electron,onida2002electronic}. The optical response and excitonic properties are subsequently evaluated using the model Bethe--Salpeter equation (mBSE) approach,~\cite{bokdam2016role,tal2020accurate,rohlfing2000electron,albrecht1998ab,rohlfing1998electron}, allowing us to determine exciton binding energies and examine the corresponding real- and momentum-space distributions. Our results reveal that chemical substitution increases the calculated exciton binding energy alongside its estimated spatial extent, whereas the structural distortion from $P4/mmm$ to $Cm$ produces comparatively modest changes in binding energy but more pronounced modifications of the excitonic distributions. The latter changes are associated with symmetry lowering and the accompanying rearrangement of the local electronic environment. This contrast points to a partial decoupling between exciton energetics and wavefunction geometry. Finally, we examine the influence of lattice polarization through an analytical phonon-screening correction to the exciton binding energies~\cite{filip2021phonon,sak1972effective} and a complementary Fr\"ohlich-polaron treatment of carrier-mass renormalization and mobility~\cite{feynman1955slow,feynman1962mobility,devreese2009frohlich}. The comparatively modest variation in the model-estimated polaron mobilities between the polymorphs further distinguishes the structural sensitivity of excitonic wavefunction geometry from that of carrier transport. Together, these results elucidate how chemical composition, structural distortion, and lattice polarization influence different aspects of the excited-state and carrier responses in imide-based defect double antiperovskites.

\section{Computational Details}

First-principles calculations based on density functional theory (DFT) and many-body perturbation theory (MBPT) were performed using the Vienna \textit{Ab initio} Simulation Package (VASP)~\cite{kresse1996efficient}. The interactions between the ionic cores and valence electrons were described using the projector-augmented-wave (PAW) method~\cite{blochl1994projector,kresse1999ultrasoft}. The exchange-correlation energy was treated within the generalized gradient approximation (GGA) using the Perdew--Burke--Ernzerhof (PBE) functional~\cite{perdew1996generalized}. A plane-wave kinetic-energy cutoff of 550~eV was employed, and the electronic self-consistency criterion was set to $10^{-8}$~eV. The atomic positions were relaxed until the Hellmann--Feynman forces on all atoms were below $10^{-3}$~eV~\AA$^{-1}$. Brillouin-zone integrations were performed using a $\Gamma$-centered $10\times10\times5$ $k$-point mesh. The optimized crystal structures were visualized using the Visualization for Electronic and Structural Analysis (VESTA) software~\cite{momma2011vesta}.

The electronic band structures were calculated using the PBE functional, including spin--orbit coupling (SOC) to account for relativistic effects, particularly those associated with the heavier Sb and Bi atoms. To examine the influence of exact exchange on the electronic structure, additional calculations were performed using the Heyd--Scuseria--Ernzerhof (HSE06) hybrid functional, incorporating 25\% short-range Hartree--Fock exchange~\cite{heyd2003hybrid,krukau2006influence}. SOC was also included in the HSE06 calculations. Quasiparticle corrections were subsequently evaluated within the single-shot $G_0W_0$ approximation, starting from the SOC-inclusive PBE electronic structure, denoted as $G_0W_0$@PBE+SOC~\cite{hedin1965new,hybertsen1986electron,shishkin2006implementation}. These calculations employed a $6\times6\times4$ $k$-point mesh and 600 electronic bands ($\mathrm{NBANDS}=600$). The convergence tests reported in the Supplementary Material (SM) assess the dielectric response with respect to the $k$-point mesh and number of bands; they should not be interpreted as a complete convergence demonstration for every $G_0W_0$ quasiparticle gap. Carrier effective masses were determined from the curvature of the SOC-inclusive electronic bands near the conduction-band minimum (CBM) and valence-band maximum (VBM) using the SUMO package~\cite{ganose2018sumo}.

The optical properties were investigated by explicitly accounting for electron--hole interactions within the mBSE approach~\cite{rohlfing1998electron,rohlfing2000electron,albrecht1998ab}. In this approach, the screened electron--hole Coulomb interaction is represented using a model dielectric function, enabling an efficient treatment of excitonic effects in the optical response~\cite{bokdam2016role,tal2020accurate}. The mBSE calculations were performed using the SOC-inclusive $G_0W_0$@PBE+SOC electronic structure. The resulting electron--hole excitation problem was solved to obtain the optical excitation energies and characterize the corresponding excitonic states. The calculations employed an $8\times8\times4$ $k$-point mesh, with 30 occupied and 30 unoccupied bands included in the electron--hole transition space. The convergence of the calculated dielectric response with respect to $k$-point sampling and the number of bands was examined for a representative structure, as shown in Figs.~S13 and S14 of the SM.

The dielectric response was further analyzed to assess the contributions of electronic and ionic polarization to excitonic screening and polaronic effects. The electronic contribution to the dielectric constant, $\varepsilon_{\infty}$, was obtained from the electronic dielectric response calculated using mBSE, whereas the ionic contribution, $\varepsilon_{\mathrm{ion}}$, was evaluated using density functional perturbation theory (DFPT)~\cite{baroni2001phonons,gonze1997dynamical}, employing a $10\times10\times5$ $k$-point mesh. These contributions were used to distinguish the high-frequency dielectric response, which accounts for electronic polarization, from the static dielectric response, which additionally includes lattice polarization.

\section{Results and discussion}

The optoelectronic response of the investigated defect-imide double antiperovskites arises from the interplay of structural characteristics, relativistic electronic effects, electron--hole interactions, and coupling to lattice polarization. We first examine the structural landscape and assess the dynamical and mechanical stability of the underlying anionic frameworks. We then investigate the influence of chemical substitution and structural polymorphism on the electronic and quasiparticle properties, followed by an analysis of the optical and excitonic responses. Finally, we examine how lattice polarization modifies exciton binding and carrier transport through Fr\"ohlich polaron formation.

\subsection{Structural landscape and lattice stability}

\subsubsection{\textbf{\textit{Crystal structure}}}

\begin{figure}[t]
	\centering
	\includegraphics[width=\columnwidth]{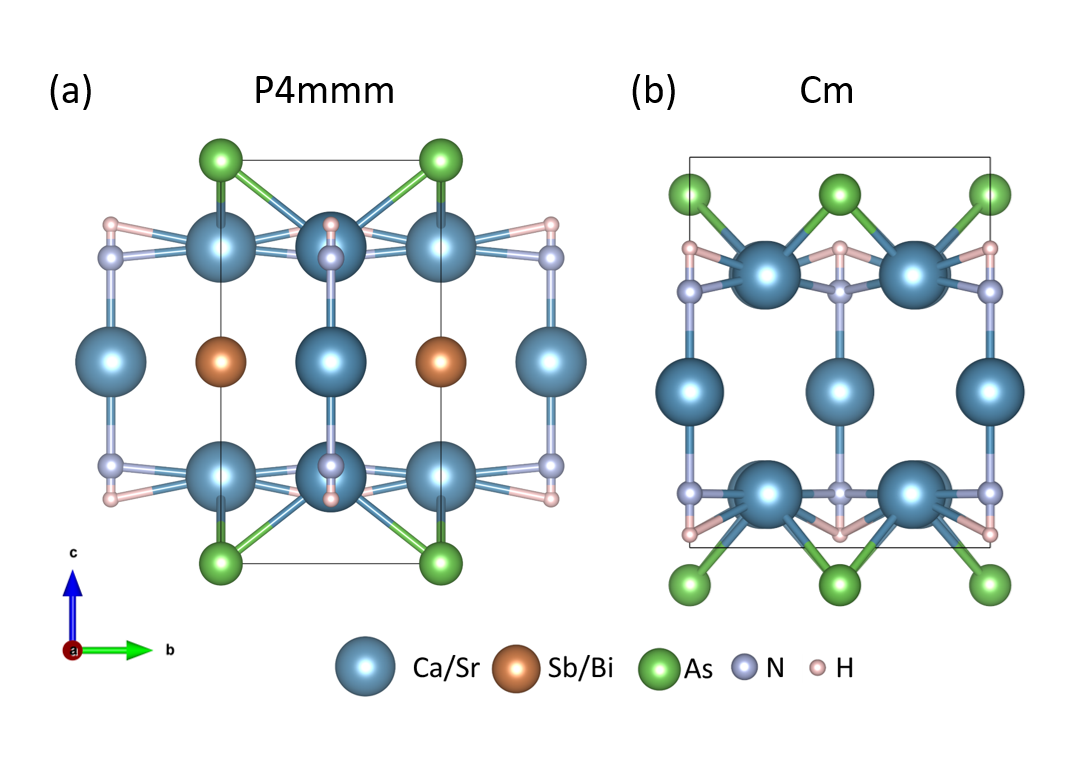}
	\caption{Crystal structures of $AE_5\mathrm{As}Pn\mathrm{(NH)}_2$:
		(a) tetragonal $P4/mmm$ and (b) monoclinic $Cm$ phases.}
	\label{fig:structures}
\end{figure}

The crystal structures of the defect-imide double antiperovskites $AE_5\mathrm{As}Pn\mathrm{(NH)}_2$ ($AE = \mathrm{Ca, Sr}$; $Pn = \mathrm{Sb, Bi}$) are shown in Fig.~\ref{fig:structures}. The investigated compounds exhibit two distinct crystallographic configurations: the tetragonal $P4/mmm$ phase and the distorted monoclinic $Cm$ phase, as illustrated in Figs.~\ref{fig:structures}(a) and \ref{fig:structures}(b), respectively~\cite{chau2025defect}. Each formula unit contains five alkaline-earth (AE) atoms, one As atom, one Pn atom, and two imide (NH) units.

In the tetragonal phase, the AE atoms form the principal cationic framework, while As and Pn occupy distinct crystallographic sites. In contrast, the monoclinic phase exhibits structural distortions that lower the crystallographic symmetry and modify the local coordination environments. In particular, the monoclinic distortion removes the inversion and fourfold rotational symmetries present in the tetragonal phase. Comparing the two polymorphs at fixed chemical composition therefore provides a basis for examining how structural distortion, together with the associated changes in crystal symmetry, lattice metrics, and local coordination, influences the electronic and excited-state properties.

\subsubsection{\textbf{\textit{Geometrical stability assessment}}}

To assess the geometrical formability of the constituent ions in $AE_5\mathrm{As}Pn\mathrm{(NH)}_2$, we evaluated the effective Goldschmidt tolerance factor ($t_{\mathrm{eff}}$) and octahedral factor ($\mu_{\mathrm{eff}}$) within an idealized hard-sphere model \cite{goldschmidt1926gesetze,shannon1976revised}. These descriptors provide an approximate measure of ionic-size compatibility within the complex antiperovskite framework. Following the effective-factor approach extended to quaternary antiperovskites \cite{han2021design}, the geometrical factors are defined as
\begin{equation}
	t_{\mathrm{eff}} =
	\frac{\left(\dfrac{r_{\mathrm{As}}+r_{\mathrm{Pn}}}{2}
		+r_{\mathrm{AE}}\right)}
	{\sqrt{2}\left(r_{\mathrm{N}}+r_{\mathrm{AE}}\right)},
	\label{eq:tolerance_factor}
\end{equation}
and
\begin{equation}
	\mu_{\mathrm{eff}} = \frac{r_{\mathrm{N}}}{r_{\mathrm{AE}}},
	\label{eq:octahedral_factor}
\end{equation}
where $r_{\mathrm{AE}}$, $r_{\mathrm{N}}$, $r_{\mathrm{As}}$, and $r_{\mathrm{Pn}}$ denote the effective ionic radii of the corresponding constituent ions. The effective ionic radii of the alkaline-earth cations (Ca and Sr) were assigned using Shannon's ionic radii \cite{shannon1976revised}, while those of N, As, Sb, and Bi were taken from reported values for the corresponding anionic species \cite{han2021design}. For the imide-containing framework, we use the ionic radius of N to represent the local anionic coordination centre in the geometrical-factor analysis. This is a coordination-centre approximation rather than an assignment of a spherical ionic radius to the entire NH$^{2-}$ molecular unit. Since the imide group is non-spherical and its effective spatial extent depends on its orientation and local bonding environment, the resulting tolerance and octahedral factors should be interpreted as approximate geometrical screening descriptors rather than quantitative measures of structural stability \cite{kieslich2015extended,dawson2021anti}.

In conventional perovskites, the octahedral factor is defined as the ratio of the B-site cation radius to the anion radius, $\mu=r_B/r_X$. Because the cationic and anionic sublattices are interchanged in antiperovskites, we use the inverse ratio, $\mu_{\mathrm{eff}}=r_X/r_B$. Correspondingly, the conventional interval $0.44<\mu<0.90$ maps to $1/0.90<\mu_{\mathrm{eff}}<1/0.44$, or approximately $1.11<\mu_{\mathrm{eff}}<2.27$. The tolerance-factor interval is retained as $0.81<t<1.11$. These reciprocal-radius considerations and the geometrical screening of antiperovskites have been discussed in the context of ion-type inversion and anion ordering \cite{han2021design}.

The effective ionic radii used in our calculations are $r_{\mathrm{Ca}} = 1.00$~\AA, $r_{\mathrm{Sr}} = 1.18$~\AA, $r_{\mathrm{N}} = 1.37$~\AA, $r_{\mathrm{As}} = 1.90$~\AA, $r_{\mathrm{Sb}} = 2.12$~\AA, and $r_{\mathrm{Bi}} = 2.21$~\AA. The calculated $t_{\mathrm{eff}}$ values (0.90--0.91) and $\mu_{\mathrm{eff}}$ values (1.16--1.37) fall within the adopted empirical ranges, indicating favourable ionic-size compatibility for all five structures. Substitution of Ca by Sr or Sb by Bi modifies these effective factors through the corresponding changes in ionic radii. The calculated lattice parameters, crystal systems, space groups, and effective factors are summarized in Table~\ref{tab:structural_properties}. Additional local coordination descriptors are provided in Table S1 of the SM.

\subsubsection{\textbf{\textit{Formation energetics}}}

To examine the energetics of compound formation, we calculated the formation energy ($E_f$) per atom of $AE_5\mathrm{As}Pn\mathrm{(NH)}_2$ with respect to the corresponding elemental reference states:
\begin{equation}
	E_f = \frac{1}{11}\left[
	E_{\mathrm{tot}} - 5E_{\mathrm{AE}} - E_{\mathrm{As}}
	- E_{\mathrm{Pn}} - 2E_{\mathrm{N}} - 2E_{\mathrm{H}}
	\right],
	\label{eq:formation_energy}
\end{equation}
where $E_{\mathrm{tot}}$ is the total energy of the compound and $E_i$ denotes the reference energy per atom of the corresponding element. All five structures exhibit negative formation energies, ranging from $-1.03$ to $-0.92$~eV/atom, as shown in Table~\ref{tab:structural_properties}. This indicates energetically favourable formation relative to the chosen elemental reference states.

To further assess the thermodynamic stability of the investigated defect-imide double antiperovskites, we calculated their decomposition energies using the PBE exchange-correlation functional with respect to selected competing binary and ternary phases (see the SM for details). All five structures exhibit positive decomposition energies, ranging from 74.75 to 835.43 meV/atom, indicating that decomposition into the selected competing phases is energetically unfavourable. The two polymorphs of Ca$_5$AsSb(NH)$_2$ exhibit decomposition energies of 834.84 and 835.43 meV/atom for the tetragonal and monoclinic phases, respectively. In comparison, Ca$_5$AsBi(NH)$_2$ and the tetragonal and monoclinic polymorphs of Sr$_5$AsBi(NH)$_2$ exhibit values of 74.75, 83.17, and 82.80 meV/atom, respectively. Together with the negative formation energies, these results support the energetic favourability of the investigated compounds relative to the elemental reference states and the selected decomposition pathways. We further examine their dynamical and mechanical stability to establish the structural stability of the proposed phases.

\begin{table*}[t]
	\caption{Crystal structures, lattice parameters, geometrical stability factors, and formation energies of the considered AE$_5$AsPn(NH)$_2$ compounds.}
	\label{tab:structural_properties}
	\centering
	\renewcommand{\arraystretch}{1.4}
	\setlength{\tabcolsep}{7pt}
	\begin{ruledtabular}
		\begin{tabular}{lccccccccc}
			Compound &
			Space group &
			Crystal system &
			\multicolumn{4}{c}{Lattice parameters} &
			$t_{\mathrm{eff}}$ &
			$\mu_{\mathrm{eff}}$ &
			$E_f$ (eV/atom) \\
			\cline{4-7}
			& & & $a$ (\AA) & $b$ (\AA) & $c$ (\AA) & $\beta$ ($^\circ$) & & & \\
			\hline
			Ca$_5$AsSb(NH)$_2$ & $P4/mmm$ & Tetragonal
			& 4.87 & 4.87 & 8.92 & 90.000 & 0.90 & 1.37 & -1.03 \\
			
			Ca$_5$AsSb(NH)$_2$ & $Cm$ & Monoclinic
			& 6.90 & 6.88 & 8.98 & 90.372 & 0.90 & 1.37 & -1.03 \\
			
			Ca$_5$AsBi(NH)$_2$ & $P4/mmm$ & Tetragonal
			& 4.92 & 4.92 & 9.00 & 90.000 & 0.91 & 1.37 & -1.00 \\
			
			Sr$_5$AsBi(NH)$_2$ & $P4/mmm$ & Tetragonal
			& 5.25 & 5.25 & 9.56 & 90.000 & 0.90 & 1.16 & -0.92 \\
			
			Sr$_5$AsBi(NH)$_2$ & $Cm$ & Monoclinic
			& 7.44 & 7.42 & 9.55 & 90.240 & 0.90 & 1.16 & -0.92 \\
		\end{tabular}
	\end{ruledtabular}
\end{table*}

\subsubsection{\textbf{\textit{Dynamical and mechanical stability}}}

The dynamical stability of the investigated compounds was examined through their phonon dispersion relations, calculated using density-functional perturbation theory (DFPT) \cite{gonze1997dynamical} with a $2\times2\times1$ supercell. The calculated phonon spectra exhibit no imaginary frequencies throughout the Brillouin zone for either the tetragonal $P4/mmm$ or monoclinic $Cm$ structures (see Fig.~S1 of the SM), indicating dynamical stability within the harmonic approximation.

The mechanical stability and elastic properties were further investigated by calculating the second-order elastic stiffness constants ($C_{ij}$) using the energy--strain approach. The calculated elastic tensors satisfy the corresponding Born stability criteria for both tetragonal and monoclinic structures \cite{mouhat2014necessary}, supporting their mechanical stability (see the SM for details). The polycrystalline bulk ($B$), shear ($G$), and Young's ($Y$) moduli, together with Poisson's ratio ($\nu$), were evaluated within the Voigt--Reuss--Hill (VRH) approximation \cite{hill1952elastic}. The calculated Pugh ratios ($B/G=1.45$--$1.50$), which are below the conventional critical value of 1.75 \cite{pugh1954xcii}, and Poisson's ratios ($\nu=0.22$--$0.23$), which are below the empirical threshold of 0.26 \cite{frantsevich1982elastic}, suggest predominantly brittle behaviour across all five structures. Furthermore, the universal elastic anisotropy index ($A^{U}=0.43$--$0.52$) \cite{ranganathan2008universal} indicates a finite degree of elastic anisotropy. To further characterize the directional dependence of the elastic response, the three-dimensional angular variations of Young's modulus, shear modulus, and Poisson's ratio were also evaluated. The corresponding representations, presented in Figs.~S2--S4 of the SM, illustrate the anisotropic elastic response of these compounds. The calculated elastic stiffness constants and the corresponding polycrystalline mechanical properties are summarized in Tables~S3 and S4, respectively.

Having established the geometrical, thermodynamic, dynamical, and mechanical characteristics of the investigated structures, we next examine how their chemical compositions and crystal structures influence the electronic states near the band edges. In particular, the presence of heavy pnictogen atoms and the distinct local coordination geometries associated with the tetragonal and monoclinic polymorphs motivate an examination of spin--orbit coupling and its influence on the electronic band structure.

\subsection{Competition between relativistic effects and quasiparticle corrections}

We first examine the influence of spin--orbit coupling (SOC) and quasiparticle self-energy corrections on the electronic structures of the investigated defect-imide double antiperovskites. To disentangle these contributions, we compare the electronic structures obtained using PBE, PBE+SOC, HSE06+SOC, and the single-shot $G_0W_0$ approximation starting from PBE+SOC wavefunctions.

At the PBE level, the tetragonal phases of Ca$_5$AsSb(NH)$_2$, Ca$_5$AsBi(NH)$_2$, and Sr$_5$AsBi(NH)$_2$ exhibit band gaps of 0.85, 0.70, and 0.57 eV, respectively. Including SOC reduces these values to 0.73, 0.31, and 0.18 eV, corresponding to gap reductions of 0.12 eV for Ca$_5$AsSb(NH)$_2$ and 0.39 eV for each of the Bi-containing compounds. This pronounced composition dependence is consistent with the stronger relativistic effects associated with the heavier Bi atom and their influence on the band-edge electronic structure. In particular, the substantial gap reduction in the Bi-containing compounds highlights the importance of including SOC when describing their electronic and optical properties.

The inclusion of nonlocal exchange through the HSE06+SOC functional increases the band gaps to 1.43, 0.93, and 0.73 eV for the three tetragonal compounds, respectively. The corresponding $G_0W_0$ quasiparticle gaps are 2.01, 1.60, and 1.42 eV. Thus, whereas SOC reduces the semilocal band gaps, the inclusion of nonlocal exchange and quasiparticle self-energy corrections increases them. These competing contributions highlight the importance of treating relativistic effects and many-body corrections consistently when estimating the electronic excitation energies of these compounds.

\begin{figure*}[t]
	\centering
	\includegraphics[width=\textwidth]{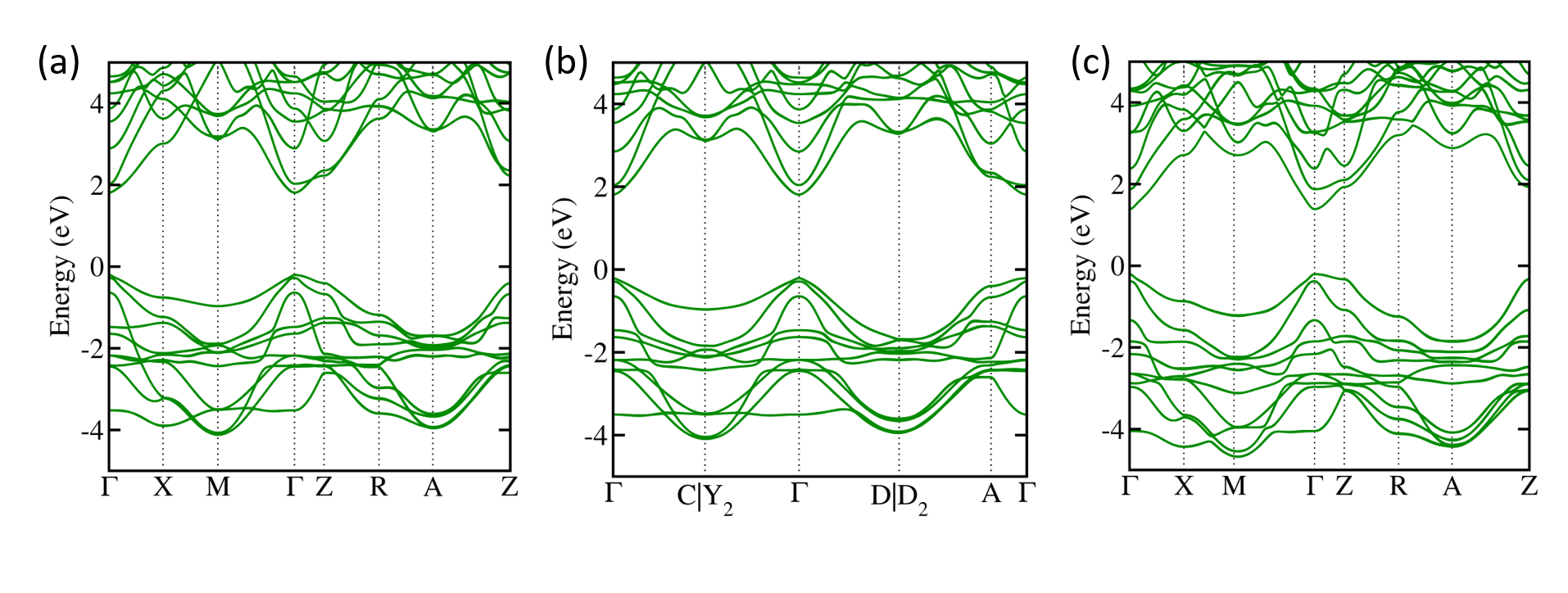}
	\caption{Quasiparticle band structures calculated using the
		$G_0W_0$@PBE+SOC approach for the tetragonal ($P4/mmm$) phases of
		(a) Ca$_5$AsSb(NH)$_2$,
		(b) Ca$_5$AsBi(NH)$_2$, and
		(c) Sr$_5$AsBi(NH)$_2$.The Fermi level is set to 0 eV}
	\label{fig:GW_bands}
\end{figure*}

The monoclinic polymorphs exhibit a similar trend, with HSE06+SOC band gaps of 1.45 and 0.76 eV for Ca$_5$AsSb(NH)$_2$ and Sr$_5$AsBi(NH)$_2$, respectively. The corresponding $G_0W_0$ calculations yield quasiparticle band gaps of 2.02 and 1.40 eV. These values are close to those obtained for the respective tetragonal polymorphs. The HSE06+SOC band structures for all five investigated structural models are provided in Fig.~S5 of the SM. The calculated $G_0W_0$ band structures of the three tetragonal compounds are presented in Fig.~\ref{fig:GW_bands}, while those of the corresponding monoclinic polymorphs are provided in Fig.~S6 of the SM. The close agreement between the quasiparticle gaps of the two polymorphs indicates that the structural distortions considered here have only a minor effect on the band gap, compared with the changes associated with chemical composition. Our standard HSE06+SOC band gaps (1.43, 0.93, and 0.73~eV for the tetragonal phases), evaluated with the default exact-exchange mixing fraction ($\alpha_{\mathrm{EXX}} = 0.25$), are smaller than the hybrid-functional values reported in Ref.~\cite{chau2025defect} (1.91, 1.33, and 1.12~eV), which employed an increased mixing fraction ($\alpha_{\mathrm{EXX}} = 0.40$). Importantly, our parameter-free $G_0W_0$@PBE+SOC quasiparticle gaps combined with mBSE excitonic corrections (Table~\ref{tab:electronic_properties}) yield optical gaps of 1.87, 1.42, and 1.22~eV, closely reproducing the experimentally measured direct optical gaps of 1.90, 1.36, and 1.16~eV~\cite{chau2025defect} without tuning the exchange parameter.

To gain further insight into the electronic structure, we examined the total density of states (TDOS) and partial density of states (PDOS) for all the investigated systems. The PDOS obtained at the PBE+SOC and $G_0W_0$@PBE+SOC level for all five structures is presented in Figs.~S7 and S8, respectively of the SM. An interesting feature is observed at the conduction-band edge of Sr$_5$AsBi(NH)$_2$. The orbital-projected electronic structures indicate a change in the dominant orbital character associated with the conduction-band minimum (CBM), from predominantly Sr-$d$ character at the PBE+SOC level to mixed Sr-$s$ and Sr-$d$ character after the $G_0W_0$@PBE+SOC quasiparticle correction. A similar change is observed for the monoclinic polymorph. In the Ca-containing compounds, the valence-band maximum (VBM) is predominantly composed of Sb/Bi-$p$ orbitals, while the CBM is mainly contributed by Ca-d orbitals. The change in the conduction-band orbital character in Sr$_5$AsBi(NH)$_2$ may be associated with a reordering of nearby conduction-band states upon inclusion of quasiparticle corrections. These orbital projections therefore provide insight into changes in the band-edge electronic structure beyond the variation in the band-gap magnitude alone.

To complement the orbital-projected electronic-structure analysis, we examined the charge distribution among the constituent atomic species using Bader charge partitioning~\cite{henkelman2006fast}. The calculated atomic charges for the tetragonal and monoclinic structures are summarized in Table~S7 of the SM. The results indicate charge transfer from the electropositive alkaline-earth atoms towards the anionic sublattice, consistent with predominantly ionic interactions. The corresponding electron localization function (ELF)~\cite{becke1990simple} distributions are presented in Fig.~S9 of the SM.

To further characterize the band-edge dispersion, we calculated the electron and hole effective masses from the curvature of the SOC-inclusive bands near the CBM and VBM. The resulting electron effective masses ($m_e^*$), hole effective masses ($m_h^*$), and reduced masses ($\mu_{\mathrm{red}}$) are summarized in Table~\ref{tab:electronic_properties}. Here, $m_e^*$ and $m_h^*$ denote the effective masses of electrons and holes, respectively, while $m_0$ represents the free-electron rest mass. The electron effective masses range from 0.221 to 0.365$m_0$, whereas the hole effective masses range from 0.407 to 0.626$m_0$ across the investigated structures. The corresponding reduced mass is defined as
\begin{equation}
	\mu_{\mathrm{red}} =
	\frac{m_e^* m_h^*}{m_e^* + m_h^*},
	\label{eq:reduced_mass}
\end{equation}
and ranges from 0.150 to 0.230$m_0$. The relatively moderate effective masses indicate appreciable band dispersion near the band edges and provide useful input for the subsequent analysis of excitonic and polaronic properties. However, effective masses alone do not determine carrier mobility, which also depends on the relevant scattering mechanisms. The directional electron and hole effective masses along the relevant high-symmetry directions, together with their harmonic-mean values, are provided in Tables~S5 and S6 of the SM for the tetragonal and monoclinic structures, respectively.

Overall, these results reveal a competition between relativistic effects and quasiparticle self-energy corrections. SOC produces a pronounced reduction of the band gap, particularly in the Bi-containing compounds, whereas HSE06 and $G_0W_0$ corrections increase the gap relative to the semilocal PBE+SOC results. The resulting quasiparticle electronic structures provide the basis for the subsequent analysis of electron--hole interactions and optical excitations.

\begin{table*}[t]
	\caption{Calculated electronic band gaps and effective masses of the investigated defect-imide double antiperovskites. The band gaps are given in eV, while the effective masses are expressed in units of the free-electron mass ($m_0$).}
	\label{tab:electronic_properties}
	\centering
	\scriptsize
	\setlength{\tabcolsep}{3pt}
	\begin{ruledtabular}
		\begin{tabular}{lcccccccc}
			Material
			& \shortstack{PBE/\\PBE+SOC}
			& \shortstack{HSE06+SOC}
			& \shortstack{$G_0W_0$@PBE+SOC}
			& \shortstack{mBSE@\\$G_0W_0$@PBE+SOC}
			& \shortstack{Exp.~\cite{chau2025defect}}
			& \shortstack{$m_e^*$($m_0$)}
			& \shortstack{$m_h^*$($m_0$)}
			& \shortstack{$\mu_{\mathrm{red}}$($m_0$)} \\
			\hline
			\shortstack[l]{Ca$_5$AsSb(NH)$_2$\\\textit{($P4/mmm$)}}
			& 0.85/0.73 & 1.43 & 2.01 & 1.87 & - & 0.365 & 0.626 & 0.230 \\
			
			\shortstack[l]{Ca$_5$AsSb(NH)$_2$\\\textit{($Cm$)}}
			& 0.85/0.73 & 1.45 & 2.02 & 1.87 & 1.90 & 0.347 & 0.594 & 0.220 \\
			
			\shortstack[l]{Ca$_5$AsBi(NH)$_2$\\\textit{($P4/mmm$)}}
			& 0.70/0.31 & 0.93 & 1.60 & 1.42 & 1.36 & 0.323 & 0.407 & 0.180 \\
			
			\shortstack[l]{Sr$_5$AsBi(NH)$_2$\\\textit{($P4/mmm$)}}
			& 0.57/0.18 & 0.73 & 1.42 & 1.22 & 1.16 & 0.221 & 0.458 & 0.150 \\
			
			\shortstack[l]{Sr$_5$AsBi(NH)$_2$\\\textit{($Cm$)}}
			& 0.57/0.18 & 0.76 & 1.40 & 1.19 & 1.14 & 0.232 & 0.446 & 0.152 \\
		\end{tabular}
	\end{ruledtabular}
	\begin{flushleft}
		\footnotesize The first and second values in the PBE/PBE+SOC column denote the band gaps calculated without and with SOC, respectively. SOC is included in the HSE06+SOC, $G_0W_0$@PBE+SOC, and mBSE@$G_0W_0$@PBE+SOC calculations. Exp. denotes the experimentally reported optical gap values~\cite{chau2025defect}. $m_e^*$ and $m_h^*$ are the electron and hole effective masses, respectively, and $\mu_{\mathrm{red}}$ is the reduced effective mass.
	\end{flushleft}
\end{table*}

\subsection{Optical response and excitonic structure}

The Coulomb attraction between photoexcited electrons and holes gives rise to excitonic optical excitations below the quasiparticle electron--hole continuum~\cite{onida2002electronic,rohlfing1998electron}. A full solution of the Bethe--Salpeter equation (BSE) is computationally demanding, particularly for the relatively dense $\mathbf{k}$-point meshes required to describe electron--hole interactions accurately~\cite{onida2002electronic,rohlfing2000electron}. We therefore employed the model Bethe--Salpeter equation (mBSE) approach to incorporate electron--hole correlations and evaluate the optical response, using the $G_0W_0$@PBE+SOC quasiparticle electronic structure as the starting point~\cite{bokdam2016role,liu2018relativistic}. In this approach, the dielectric screening entering the electron--hole interaction kernel is described using a model function of the form~\cite{bokdam2016role,cappellini1993model}
\begin{equation}
	f(x) = 1 - (1-0.088)
	\exp\left(-\frac{x^2}{4z^2}\right),
	\label{eq:mbse_fitting}
\end{equation}
where $x$ denotes the magnitude of the momentum transfer and $z$ is a fitting parameter that determines the momentum dependence of the model screening function. The parameters of this function were obtained by fitting the model to the calculated $G_0W_0$ dielectric functions for each of the five investigated structural models. The corresponding dielectric functions and model-fitting curves are presented in Fig.~S10 of the SM. The fitted model dielectric functions were subsequently used to construct the screened electron--hole interaction kernel in the mBSE calculations. The quasiparticle corrections to the electronic band structure are incorporated through a scissor operator, which rigidly shifts the conduction bands relative to the valence bands to reproduce the corresponding $G_0W_0$ quasiparticle band gap. This approach enables the evaluation of electron--hole interactions and the resulting excitonic optical response at substantially reduced computational cost compared with a full BSE calculation~\cite{bokdam2016role}.

\begin{figure*}[t]
	\centering
	\includegraphics[width=\textwidth]{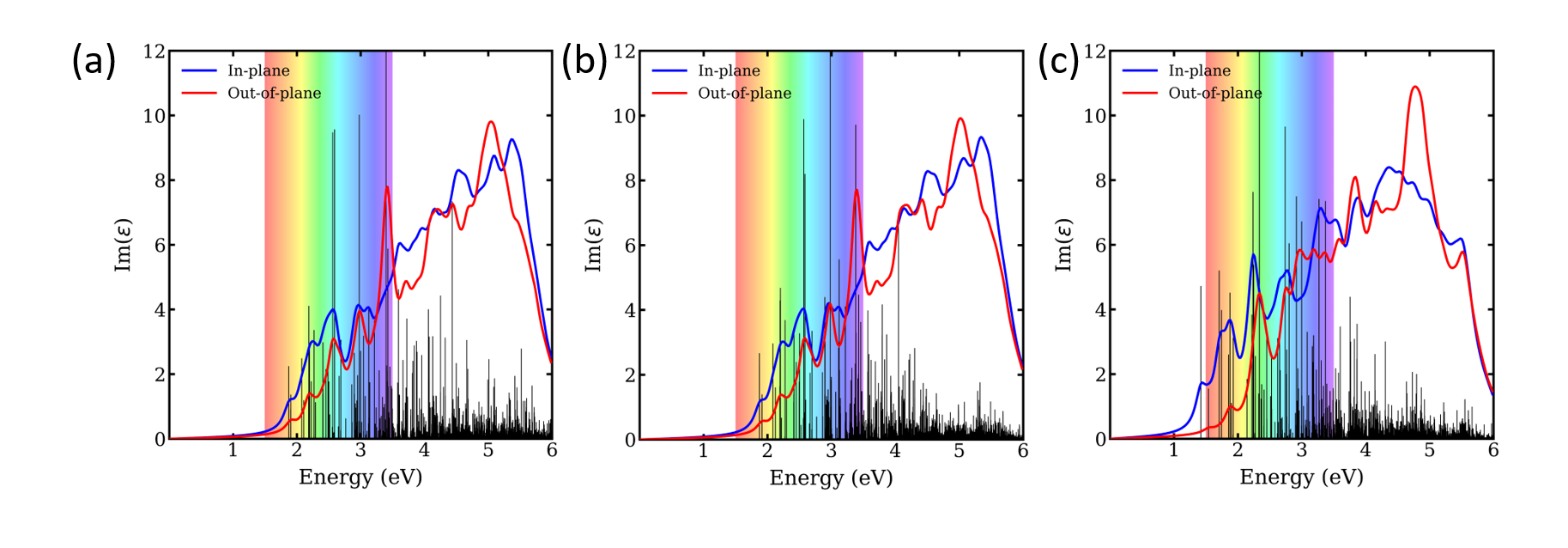}
	\caption{Calculated imaginary part of the frequency-dependent
		dielectric function, $\varepsilon_2(\omega)$, for the tetragonal
		$P4/mmm$ phases of (a) Ca$_5$AsSb(NH)$_2$,
		(b) Ca$_5$AsBi(NH)$_2$, and (c) Sr$_5$AsBi(NH)$_2$.
		The blue and red curves represent the in-plane and out-of-plane
		optical responses, respectively. The black vertical lines
		indicate the oscillator strengths of the individual optical
		transitions. The shaded region represents the visible-light
		spectral range.}
	\label{fig:optical_spectrum}
\end{figure*}

The macroscopic optical response is characterized by the frequency-dependent complex dielectric function~\cite{onida2002electronic},
\begin{equation}
	\varepsilon(\omega) =
	\varepsilon_1(\omega) + i\varepsilon_2(\omega).
	\label{eq:dielectric_function}
\end{equation}
The imaginary part, $\varepsilon_2(\omega)$, describes the optical absorption spectrum and is presented for the tetragonal $P4/mmm$ phases in Fig.~\ref{fig:optical_spectrum}; the corresponding spectra for the monoclinic $Cm$ phases are provided in Fig.~S11 of the SM. The real parts, $\varepsilon_1(\omega)$, of the frequency-dependent dielectric functions for all five investigated structural models are presented in Fig.~S12 of the SM. The calculated spectra exhibit pronounced anisotropy between the in-plane and out-of-plane optical responses across the visible and ultraviolet regions. This directional dependence is particularly evident in Sr$_5$AsBi(NH)$_2$, where the spectral peak positions and intensities differ appreciably with the polarization direction of the incident electric field. These differences reflect the influence of crystal structure and chemical composition on the optical response.

The vertical lines in Fig.~\ref{fig:optical_spectrum} indicate the oscillator strengths associated with the individual optical excitations. Their distribution shows that the calculated optical response receives contributions from multiple excitonic transitions, rather than being dominated throughout the spectrum by a single excitation. The resulting spectral features therefore reflect the combined contributions of the excitonic states and their respective oscillator strengths.

To further characterize the optical absorption, we calculated the absorption coefficient, $\alpha(\omega)$, from the dielectric function using
\begin{equation}
	\alpha(\omega) =
	\frac{\sqrt{2}\,\omega}{c}
	\left[
	\sqrt{\varepsilon_1^2(\omega)+\varepsilon_2^2(\omega)}
	-\varepsilon_1(\omega)
	\right]^{1/2},
	\label{eq:absorption_coefficient}
\end{equation}
where $c$ is the speed of light in vacuum. The resulting spectra, shown in Fig.~S15 of the SM, exhibit pronounced absorption onsets and a systematic redshift upon substituting Sb with Bi. This trend is consistent with the reduction in the calculated quasiparticle band gaps, together with the contribution of electron--hole interactions to the optical excitation energies. The absorption coefficients increase rapidly with photon energy, reaching values of approximately $10^6$~cm$^{-1}$, indicative of strong optical absorption in the investigated spectral range.

To verify whether the photovoltaic potential predicted at the single-particle level~\cite{chau2025defect} persists when excitonic effects are explicitly included, we evaluated the spectroscopic-limited maximum efficiency (SLME)~\cite{yu2012identification,shockley1991detailed} at 293.15~K under the AM1.5G spectrum using our mBSE@$G_0W_0$@PBE+SOC absorption coefficients and quasiparticle band gaps. Squared transition dipole matrix elements ($P^2$) confirm that the VBM-to-CBM transitions at $\Gamma$ are dipole allowed across all five structural models [Figs.~S16(a)--(e) of the SM]. The resulting saturation SLME values at 100~$\mu$m thickness [Fig.~S16(f) of the SM] reach 22.17\% and 22.15\% for tetragonal and monoclinic Ca$_5$AsSb(NH)$_2$, 30.20\% for tetragonal Ca$_5$AsBi(NH)$_2$, and 32.50\% and 32.85\% for tetragonal and monoclinic Sr$_5$AsBi(NH)$_2$. These excitonic-level efficiencies closely corroborate the high photovoltaic potential reported in Ref.~\cite{chau2025defect}, while confirming that the tetragonal-to-monoclinic distortion has a negligible impact on the macroscopic efficiency limit.


To investigate the microscopic characteristics of the excitonic states, we first evaluate their binding energies and characteristic spatial parameters. The exciton binding energy ($E_{\mathrm{B}}$) is defined as the energy difference between the $G_0W_0$@PBE+SOC quasiparticle gap and the corresponding mBSE excitation energy:
\begin{equation}
	E_{\mathrm{B}} =
	E_g^{G_0W_0} - E_{\mathrm{exc}}^{\mathrm{mBSE}},
	\label{eq:exciton_binding_energy}
\end{equation}
where $E_{\mathrm{exc}}^{\mathrm{mBSE}}$ denotes the excitation energy of the lowest-energy optically allowed (bright) excitonic state obtained from the mBSE calculation. In evaluating this difference, we use the direct $G_0W_0$ quasiparticle gap at the same transition wave vector (here, $\Gamma$), rather than an indirect fundamental gap. The calculated binding energies are 0.14, 0.15, 0.18, 0.20, and 0.21 eV for Ca$_5$AsSb(NH)$_2$ ($P4/mmm$), Ca$_5$AsSb(NH)$_2$ ($Cm$), Ca$_5$AsBi(NH)$_2$ ($P4/mmm$), Sr$_5$AsBi(NH)$_2$ ($P4/mmm$), and Sr$_5$AsBi(NH)$_2$ ($Cm$), respectively. These values exceed the thermal energy at room temperature ($k_{\mathrm{B}}T \sim 26$ meV), indicating that excitonic effects can remain relevant under ambient conditions. The corresponding characteristic excitonic temperatures are defined as
\begin{equation}
	T_{\mathrm{exc}} = \frac{E_{\mathrm{B}}}{k_{\mathrm{B}}},
	\label{eq:exciton_temperature}
\end{equation}
where $k_{\mathrm{B}}$ is the Boltzmann constant. $T_{\mathrm{exc}}$ represents the temperature corresponding to the exciton binding-energy scale, expressed through $E_{\mathrm{B}}=k_{\mathrm{B}}T_{\mathrm{exc}}$. It is an energy-equivalent characteristic scale rather than a literal exciton dissociation temperature; the actual thermal stability of an exciton also depends on screening, disorder, and other material-specific effects.

Within the effective-mass approximation, the characteristic exciton radius is estimated using
\begin{equation}
	r_{\mathrm{exc}} =
	\frac{m_0}{\mu_{\mathrm{red}}}\,
	\varepsilon_{\mathrm{eff}}\,a_0,
	\label{eq:exciton_radius}
\end{equation}
where $m_0$ is the free-electron mass, $\mu_{\mathrm{red}}$ is the reduced effective mass, $\varepsilon_{\mathrm{eff}}$ is the effective dielectric constant, and $a_0$ is the Bohr radius. In the present calculation, $\varepsilon_{\mathrm{eff}}$ is taken as the electronic dielectric constant $\varepsilon_{\infty}$. The electron--hole probability density at zero separation for a hydrogenic $nS$ exciton is given by
\begin{equation}
	|\phi_n(0)|^2 =
	\frac{1}{\pi n^3 r_{\mathrm{exc}}^3}.
	\label{eq:exciton_probability}
\end{equation}
Using electronic dielectric screening and the ground-state value $n=1$, the estimated exciton radii range from 1.09 to 2.00 nm, while the corresponding electron--hole probability densities at zero separation range from $0.40$ to $2.46\times10^{26}$~m$^{-3}$. The increase in the estimated exciton radius upon substituting Sb with Bi and Ca with Sr is accompanied by a reduction in the contact probability, indicating a more spatially extended electron--hole envelope within the hydrogenic approximation. The smaller contact probabilities in the Sr-containing compounds may be associated with reduced electron--hole overlap. However, the radiative recombination rate also depends on the relevant optical transition matrix elements; therefore, the contact probability alone does not establish a quantitative lifetime trend. The formalism used to estimate exciton radiative lifetimes is provided in the SM. The calculated excitonic parameters for all five structures are summarized in Table~\ref{tab:exciton_properties}.

\begin{table*}[t]
	\caption{Calculated excitonic parameters of the investigated
		compounds. $E_{\mathrm{B}}$ is the exciton binding energy,
		$T_{\mathrm{exc}}=E_{\mathrm{B}}/k_{\mathrm{B}}$ is the characteristic
		temperature associated with the binding-energy scale, $r_{\mathrm{exc}}$
		is the hydrogenic estimate of the exciton radius, and
		$|\phi_1(0)|^2$ is the corresponding electron--hole probability
		density at zero separation. $\Delta E_{\mathrm{B}}^{\mathrm{ph}}$
		denotes the phonon-induced correction, and
		$E_{\mathrm{B}}^{\mathrm{ph}}$ is the corresponding
		phonon-corrected binding energy.}
	\label{tab:exciton_properties}
	\centering
	\small
	\setlength{\tabcolsep}{5pt}
	\begin{ruledtabular}
		\begin{tabular}{lccccccc}
			Material
			& $E_{\mathrm{B}}$ (eV)
			& $T_{\mathrm{exc}}$ (K)
			& $r_{\mathrm{exc}}$ (nm)
			& $|\phi_1(0)|^2$ ($10^{26}$ m$^{-3}$)
			& $\Delta E_{\mathrm{B}}^{\mathrm{ph}}$ (meV)
			& Reduction (\%)
			& $E_{\mathrm{B}}^{\mathrm{ph}}$ (meV) \\
			\hline
			\makecell[l]{Ca$_5$AsSb(NH)$_2$\\($P4/mmm$)}
			& 0.14 & 1624 & 1.09 & 2.46 & $-22.06$ & 15.76 & 117.94 \\
			
			\makecell[l]{Ca$_5$AsSb(NH)$_2$\\($Cm$)}
			& 0.15 & 1740 & 1.14 & 2.15 & $-21.90$ & 14.60 & 128.12 \\
			
			\makecell[l]{Ca$_5$AsBi(NH)$_2$\\($P4/mmm$)}
			& 0.18 & 2088 & 1.63 & 0.73 & $-21.00$ & 11.66 & 159.01 \\
			
			\makecell[l]{Sr$_5$AsBi(NH)$_2$\\($P4/mmm$)}
			& 0.20 & 2320 & 2.00 & 0.40 & $-19.62$ & 9.81 & 180.38 \\
			
			\makecell[l]{Sr$_5$AsBi(NH)$_2$\\($Cm$)}
			& 0.21 & 2436 & 2.00 & 0.40 & $-19.63$ & 9.35 & 190.36 \\
		\end{tabular}
	\end{ruledtabular}
\end{table*}

Complementary Wannier--Mott estimates obtained using electronic and static dielectric screening are provided in Table~S8 of the SM. The exciton binding energies reported previously for these compositions (28--48~meV)~\cite{chau2025defect} lie between the upper and lower Wannier--Mott estimates obtained using $\varepsilon_\infty$ and $\varepsilon_{\mathrm{s}}$, respectively. Our mBSE calculations yield higher binding energies of 140--210~meV; in particular, the mBSE values for the Bi-containing compounds exceed the corresponding upper Wannier--Mott estimates. This comparison highlights the sensitivity of the estimated binding energies to the treatment of electron--hole interactions and dielectric screening. Importantly, the mBSE eigenvectors also provide access to the multiband composition and real- and momentum-space distributions of the optically active excitonic states, which are not resolved by hydrogenic effective-mass estimates.

\begin{figure*}[t]
	\centering
	\includegraphics[width=\textwidth]{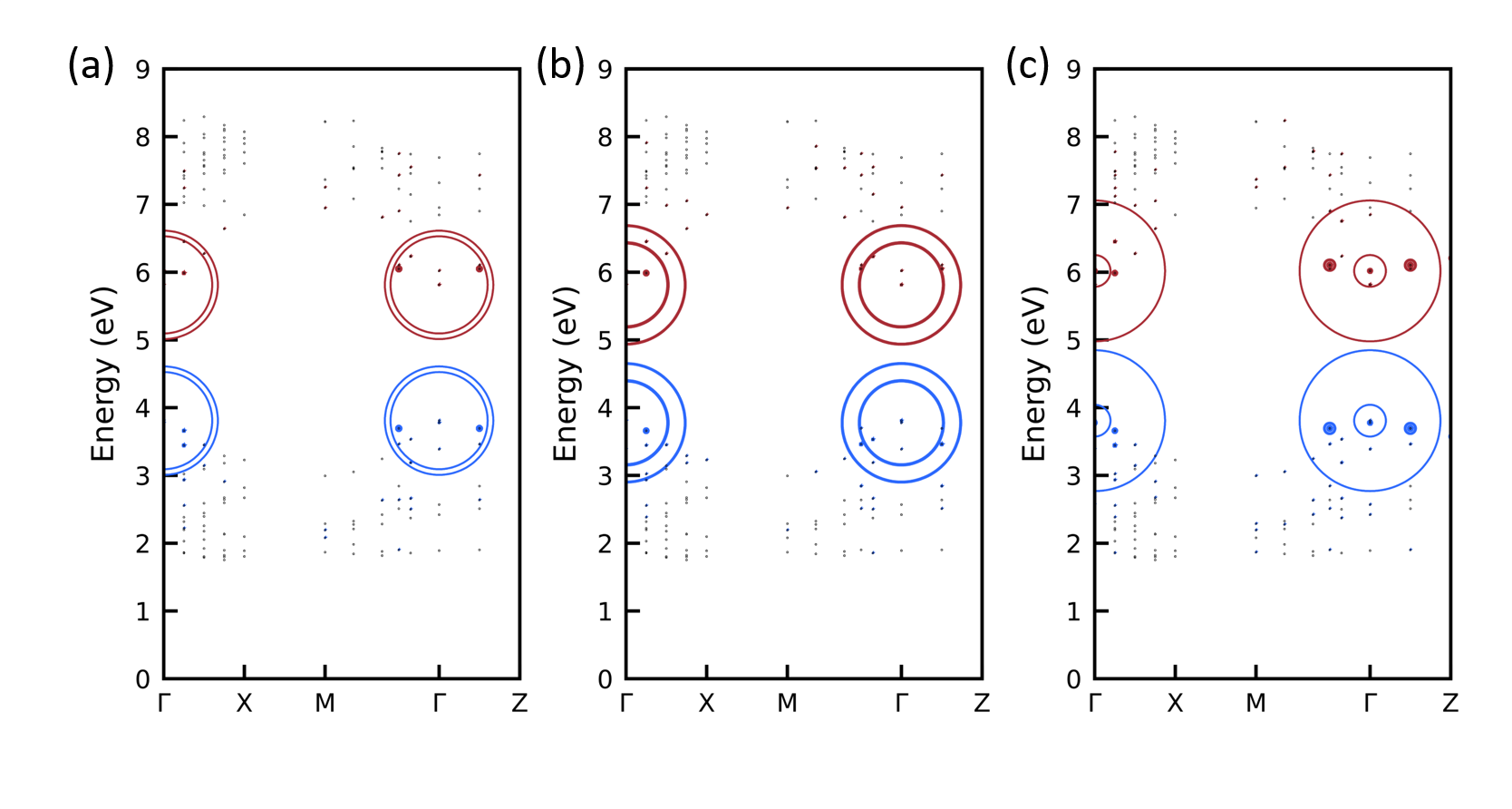}
	\caption{BSE fat-band representations of the first three
		excitonic transitions of Ca$_5$AsSb(NH)$_2$ in the
		tetragonal $P4/mmm$ phase: (a)--(c) the first, second, and
		third transitions, respectively. The vertical axis represents
		the energy eigenvalues, and the horizontal axis follows the
		$\Gamma$--X--M--$\Gamma$--Z high-symmetry path. Blue and red
		circles indicate the valence- and conduction-band-related
		states, respectively. The circle radii represent the magnitudes
		of the corresponding electron--hole coupling coefficients,
		while the small dots indicate the underlying electronic states.}
	\label{fig:bse_fatband}
\end{figure*}

To identify the quasiparticle states contributing to the bound electron--hole pairs, we projected the mBSE eigenvectors onto the underlying electronic band structures using a fat-band representation. The first three optically allowed excitonic transitions for each structural model are presented in Fig.~\ref{fig:bse_fatband} and Figs.~S17--S20 of the SM. The blue and red markers represent the valence- and conduction-band contributions, respectively, with their radii scaled according to the magnitudes of the corresponding mBSE eigenvector coefficients. For all five structures, the dominant transition weights are concentrated predominantly around the $\Gamma$ point, indicating that the optically active excitonic states are primarily composed of interband transitions in the vicinity of the band edges. The distribution of these weights reveals contributions from multiple valence and conduction bands, demonstrating the multiband character of the excitonic states rather than a description in terms of a single isolated band-to-band transition.

Although the dominant momentum-space contributions remain centred around $\Gamma$ across all investigated structures, their relative distributions vary with chemical composition and crystal structure. The tetragonal $P4/mmm$ structure has inversion symmetry, so its electronic states can be classified by parity at inversion-invariant wave vectors. In the monoclinic $Cm$ structure, inversion is absent and parity is no longer a good quantum number. The reduction from point group $D_{4h}$ to $C_s$ changes the symmetry constraints on state mixing and can redistribute the relative contributions of valence- and conduction-band transitions to the excitonic eigenvectors. The differences between the fat-band distributions of the two polymorphs are consistent with this symmetry lowering, although the accompanying changes in local coordination and band dispersion also contribute. The finite transition matrix elements calculated at $\Gamma$ for all five structures establish that the corresponding band-edge transitions are optically allowed; the absence of inversion in $Cm$ should not, by itself, be interpreted as evidence of enhanced optical activity. Substitution of Sb with Bi and Ca with Sr also modifies the participating electronic states while retaining the predominantly $\Gamma$-centred character of the excitonic transitions.

\begin{figure}[t]
	\centering
	\includegraphics[width=\columnwidth]{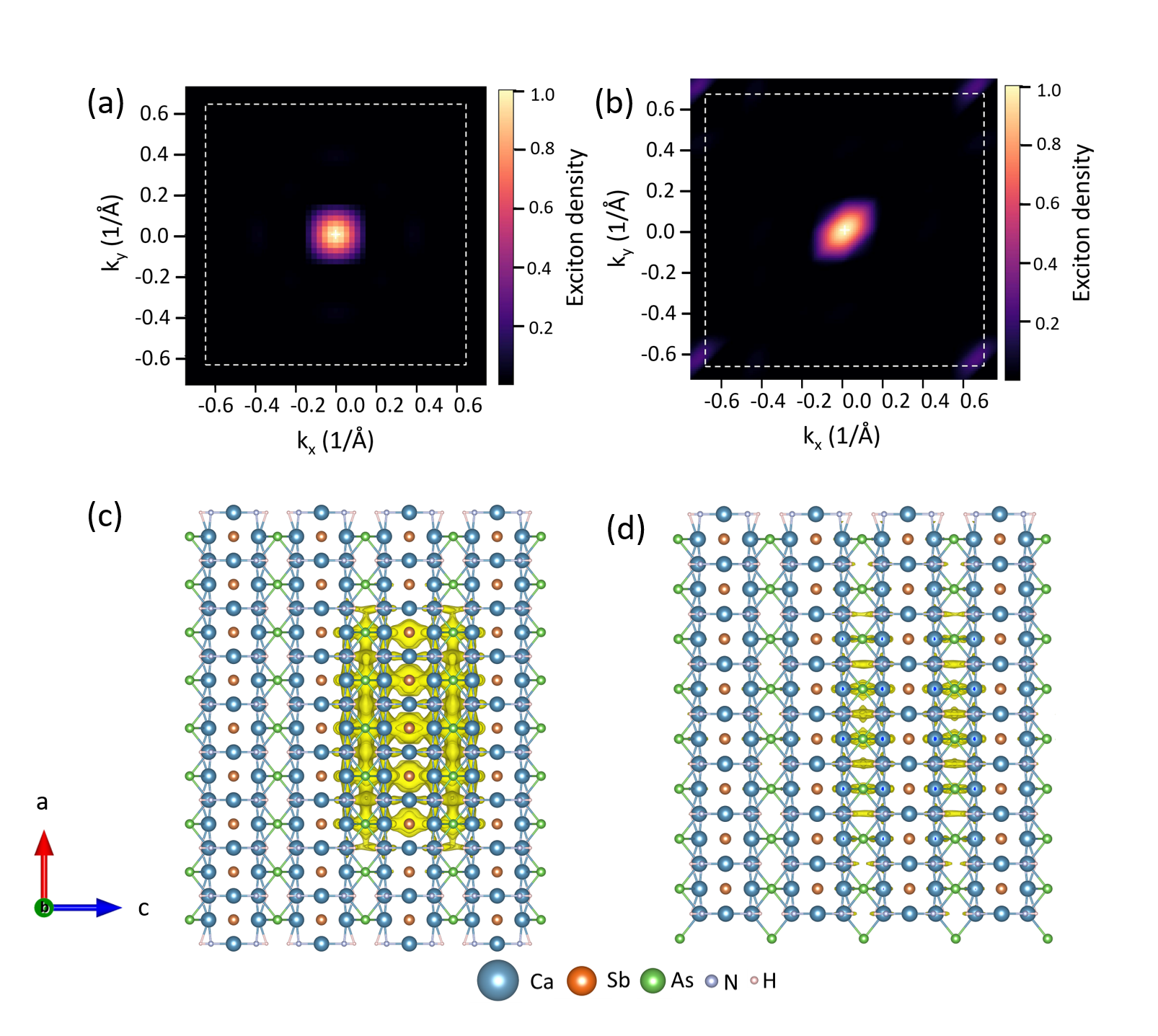}
	\caption{Momentum-space exciton densities and real-space
		exciton wavefunctions of Ca$_5$AsSb(NH)$_2$ in the
		tetragonal $P4/mmm$ and monoclinic $Cm$ phases.
		Panels (a) and (b) show the exciton density distributions
		within the first Brillouin zone for the tetragonal and
		monoclinic phases, respectively, in the $k_x$--$k_y$ plane.
		Panels (c) and (d) show the corresponding real-space exciton
		wavefunctions, with the hole fixed at an Sb atom. The
		isosurfaces are plotted at 5\% of the maximum exciton
		probability density.}
	\label{fig:exciton_density}
\end{figure}

The corresponding momentum-space distributions provide a complementary description of these excitations. The $\mathbf{k}$-resolved mBSE weight for an excitonic state $S$ is defined as
\begin{equation}
	\rho_{\mathrm{exc}}^{S}(\mathbf{k}) =
	\sum_{v,c}
	\left|A_{vck}^{S}\right|^2,
	\label{eq:exciton_k_density}
\end{equation}
where $A_{vck}^{S}$ denotes the mBSE eigenvector coefficient associated with the transition from valence band $v$ to conduction band $c$ at wave vector $\mathbf{k}$. Thus, $\rho_{\mathrm{exc}}^{S}(\mathbf{k})$ quantifies the contribution of each wave vector to the excitonic state after summing over the included valence- and conduction-band transitions. The momentum-space distributions for the tetragonal and monoclinic phases of Ca$_5$AsSb(NH)$_2$ are shown in Figs.~\ref{fig:exciton_density}(a) and \ref{fig:exciton_density}(b), respectively. The tetragonal structure exhibits a compact distribution centred around $\Gamma$, with an approximately isotropic profile in the plotted plane. In contrast, the monoclinic phase displays a more anisotropic distribution, characterized by an elongated and tilted profile. This difference is consistent with the lower rotational symmetry of the $Cm$ structure and the accompanying changes in its electronic dispersion and local environment. The structural distortion therefore modifies the relative wave-vector contributions to the excitonic eigenvector, even though the dominant weight remains concentrated around $\Gamma$. For comparison, the reciprocal-space exciton-density distributions for Ca$_5$AsBi(NH)$_2$ ($P4/mmm$) and the tetragonal and monoclinic phases of Sr$_5$AsBi(NH)$_2$ are presented in Fig.~S21 of the SM. The tetragonal Ca$_5$AsBi(NH)$_2$ and Sr$_5$AsBi(NH)$_2$ structures retain relatively compact distributions around $\Gamma$, whereas the monoclinic Sr$_5$AsBi(NH)$_2$ structure also exhibits a modified, anisotropic momentum-space distribution.

To examine the corresponding real-space structure of the excitons, we visualized the electron probability density with the hole fixed at a pnictogen site (Sb or Bi), where the dominant VBM charge density is localized. The VBM and CBM charge densities used to identify the band-edge localization are shown in Fig.~S22 of the SM. The corresponding excitonic electron distributions for the tetragonal and monoclinic polymorphs of Ca$_5$AsSb(NH)$_2$ are shown in Figs.~\ref{fig:exciton_density}(c) and \ref{fig:exciton_density}(d), respectively, while those for Ca$_5$AsBi(NH)$_2$ ($P4/mmm$) and the two polymorphs of Sr$_5$AsBi(NH)$_2$ are presented in Fig.~S23 of the SM. At an isosurface level of 5\% of the maximum probability density, all five structures exhibit electron distributions concentrated around the fixed hole, with appreciable extension over neighbouring atomic sites. Nevertheless, the spatial characteristics vary with composition and crystal structure. For Ca$_5$AsSb(NH)$_2$, the tetragonal phase exhibits a relatively continuous electron distribution, whereas the monoclinic phase displays a more fragmented and spatially modulated pattern. Upon replacing Sb with Bi, the electron distribution becomes more extended along the crystallographic $c$ direction. A similar anisotropic and elongated distribution is observed in both polymorphs of Sr$_5$AsBi(NH)$_2$, although differences in their spatial patterns remain evident.

Taken together, the real- and momentum-space analyses reveal a partial decoupling between exciton energetics and wavefunction geometry. Across the investigated compositions, the exciton binding energies increase from 0.14 to 0.20 eV for the tetragonal structures, accompanied by an increase in the hydrogenic estimates of the characteristic exciton radius. For the two compositions with both tetragonal and monoclinic polymorphs, the structural distortion produces only modest changes in the binding energies and estimated radii. In contrast, the momentum- and real-space distributions exhibit appreciable modifications between the two polymorphs. These results indicate that the energetic characteristics of the excitons are comparatively robust against the structural distortions considered here, whereas their wavefunction geometry is more sensitive to the symmetry lowering and accompanying changes in local coordination and electronic dispersion. We therefore identify a symmetry-associated reshaping of exciton wavefunctions, rather than a symmetry-only effect isolated from the other structural changes.

\subsection{The dual role of lattice polarization: Phonon-screened excitons and Fr\"ohlich polarons}

In addition to the electronic screening captured by the mBSE framework, the polar lattice response provides an additional screening channel for the electron--hole interaction. In polar semiconductors, the ionic contribution to the dielectric response, mediated by lattice vibrations, can reduce the exciton binding energy. To estimate this contribution, we evaluated the phonon-induced correction to the exciton binding energy using an analytical Fr\"ohlich model~\cite{sak1972effective,filip2021phonon}. Within this model, assuming isotropic parabolic electronic bands in the long-wavelength ($q\rightarrow 0$) limit, the lattice-polarization correction is expressed as
\begin{equation}
	\Delta E_{\mathrm{B}}^{\mathrm{ph}}
	=
	-2\hbar\omega_{\mathrm{LO}}
	\left(1-\frac{\varepsilon_\infty}{\varepsilon_{\mathrm{s}}}\right)
	\frac{
		\sqrt{1+\dfrac{\hbar\omega_{\mathrm{LO}}}{E_{\mathrm{B}}}}+3
	}{
		\left(1+\sqrt{1+\dfrac{\hbar\omega_{\mathrm{LO}}}{E_{\mathrm{B}}}}\right)^3
	},
	\label{eq:phonon_screening}
\end{equation}
where $\omega_{\mathrm{LO}}$ is the characteristic longitudinal optical (LO) phonon angular frequency, $\varepsilon_\infty$ and $\varepsilon_{\mathrm{s}}$ are the high-frequency and static dielectric constants, respectively, and $E_{\mathrm{B}}$ is the uncorrected mBSE exciton binding energy. The binding energy including this analytical phonon-screening correction is estimated as
\begin{equation}
	E_{\mathrm{B}}^{\mathrm{ph}} =
	E_{\mathrm{B}}+\Delta E_{\mathrm{B}}^{\mathrm{ph}}.
	\label{eq:phonon_corrected_binding}
\end{equation}
The negative sign of $\Delta E_{\mathrm{B}}^{\mathrm{ph}}$ reflects the reduction of the effective electron--hole attraction due to the additional screening associated with lattice polarization. As summarized in Table~\ref{tab:exciton_properties}, the calculated corrections range from $-19.62$ to $-22.06$ meV across the five investigated structures. Relative to the uncorrected binding energies, these values correspond to reductions of approximately 9.35--15.76\%, yielding corrected binding energies of about 118--190 meV. Thus, within the assumptions of the analytical model, lattice polarization provides a non-negligible screening contribution, while the excitonic states retain appreciable binding after this correction is included.

\begin{table*}[t]
	\caption{Calculated electron-polaron parameters of the
		investigated defect-imide double antiperovskites at room temperature.
		$f_{\mathrm{LO}}$ is the characteristic LO-phonon
		frequency, $\alpha$ is the Fr\"ohlich coupling parameter,
		$R_{\mathrm{f}}$ is the polaron radius, $|E_{\mathrm{p}}|$
		is the magnitude of the polaron energy lowering,
		$m_{\mathrm{p}}/m^*$ is the polaron-to-band effective-mass
		ratio, and $\mu_{\mathrm{p}}$ is the electron-polaron
		mobility.}
	\label{tab:polaron}
	\centering
	\small
	\setlength{\tabcolsep}{7pt}
	\begin{ruledtabular}
		\begin{tabular}{lcccccc}
			Material (phase)
			& $f_{\mathrm{LO}}$ (THz)
			& $\alpha$
			& $R_{\mathrm{f}}$ ({\AA})
			& $|E_{\mathrm{p}}|$ (meV)
			& $m_{\mathrm{p}}/m^*$
			& $\mu_{\mathrm{p}}$
			(cm$^2$ V$^{-1}$ s$^{-1}$) \\
			\hline
			Ca$_5$AsSb(NH)$_2$ ($P4/mmm$)
			& 8.556 & 1.80 & 84.40 & 65.10 & 1.54 & 39.36 \\
			Ca$_5$AsSb(NH)$_2$ ($Cm$)
			& 8.309 & 1.79 & 82.72 & 62.86 & 1.53 & 41.68 \\
			Ca$_5$AsBi(NH)$_2$ ($P4/mmm$)
			& 8.020 & 1.45 & 89.65 & 48.95 & 1.40 & 59.46 \\
			Sr$_5$AsBi(NH)$_2$ ($P4/mmm$)
			& 7.012 & 1.32 & 85.00 & 38.90 & 1.33 & 99.50 \\
			Sr$_5$AsBi(NH)$_2$ ($Cm$)
			& 7.018 & 1.33 & 84.53 & 39.23 & 1.34 & 93.36 \\
		\end{tabular}
	\end{ruledtabular}
\end{table*}

Lattice polarization also influences the properties of individual charge carriers. The coupling of electrons and holes to LO phonons can give rise to Fr\"ohlich polarons, in which the carrier is dressed by the lattice-polarization field, thereby modifying its energy and effective mass and influencing its transport properties~\cite{feynman1955slow,devreese2009frohlich}. We examine this complementary effect using the Feynman--Hellwarth polaron model. To characterize the carrier--phonon interaction in the defect-imide double antiperovskites, we evaluated the Fr\"ohlich coupling parameter, polaron radius, formation energy, effective-mass enhancement, and carrier mobility.

The strength of the long-range carrier--phonon interaction is characterized by the dimensionless Fr\"ohlich coupling parameter $\alpha$, defined as~\cite{devreese2009frohlich}
\begin{equation}
	\alpha =
	\frac{e^2}{4\pi\varepsilon_0\hbar}
	\left(
	\frac{1}{\varepsilon_\infty}
	-
	\frac{1}{\varepsilon_{\mathrm{s}}}
	\right)
	\left(
	\frac{m^*}{2\hbar\omega_{\mathrm{LO}}}
	\right)^{1/2},
	\label{eq:frohlich_alpha}
\end{equation}
where $e$ is the elementary charge, $\varepsilon_0$ is the vacuum permittivity, $m^*$ is the carrier band effective mass, and $\omega_{\mathrm{LO}}$ is the characteristic LO-phonon angular frequency. The corresponding phonon frequencies are expressed as $\omega_{\mathrm{LO}}=2\pi f_{\mathrm{LO}}$, where $f_{\mathrm{LO}}$ is the frequency in cycles per second. Since these compounds exhibit multiple infrared-active optical phonon branches, an effective LO-phonon frequency was obtained using the Hellwarth thermal-averaging scheme~\cite{hellwarth1999mobility}, based on the phonon frequencies calculated using density-functional perturbation theory~\cite{baroni2001phonons,gonze1997dynamical}. The resulting characteristic frequencies range from 7.012 to 8.556 THz, while the corresponding Fr\"ohlich coupling parameters range from 1.32 to 1.80 (Table~\ref{tab:polaron}). These values indicate weak-to-intermediate carrier--phonon coupling within the Fr\"ohlich model.

The carrier--lattice interaction also lowers the energy of the coupled polaronic state relative to the corresponding bare-carrier state. Within the analytical approximation employed here, the polaron formation energy is estimated as~\cite{feynman1955slow,devreese2009frohlich}
\begin{equation}
	E_{\mathrm{p}} =
	-\left(\alpha+0.0123\alpha^2\right)
	\hbar\omega_{\mathrm{LO}}.
	\label{eq:polaron_formation}
\end{equation}
The magnitude of this energy provides an estimate of the energetic stabilization associated with polaron formation. The spatial extent of the lattice-polarization cloud is characterized by the polaron radius $R_{\mathrm{f}}$. The calculated radii range from 82.7 to 89.7 {\AA}, indicating spatially extended Fr\"ohlich polarons rather than carriers localized within individual atomic sites.

Polaron formation also modifies the carrier effective mass through the dynamical response of the lattice. In the weak-coupling limit, the mass enhancement can be expressed as~\cite{devreese2009frohlich}
\begin{equation}
	m_{\mathrm{p}} =
	m^*
	\left(
	1+\frac{\alpha}{6}
	+\frac{\alpha^2}{40}
	+\cdots
	\right),
	\label{eq:polaron_mass}
\end{equation}
where $m^*$ and $m_{\mathrm{p}}$ are the bare band mass and polaron effective mass, respectively. Equation~\eqref{eq:polaron_mass} displays the leading terms of the weak-coupling expansion, whereas the mass ratios reported in Table~\ref{tab:polaron} were evaluated numerically including higher-order contributions beyond the terms explicitly shown. The calculated ratios indicate a moderate enhancement, with $m_{\mathrm{p}}/m^*$ ranging from 1.33 to 1.54. Thus, carrier--phonon coupling renormalizes the band masses without producing an extreme mass enhancement in the investigated structures.

To estimate the resulting carrier transport, we calculated the Fr\"ohlich-polaron mobility using the Feynman--Hellwarth model~\cite{feynman1962mobility,hellwarth1999mobility}:
\begin{equation}
	\mu_{\mathrm{p}} =
	\frac{3\sqrt{\pi}e}{\omega_{\mathrm{LO}}m^*\alpha}
	\frac{\sinh(\beta/2)}{\beta^{5/2}}
	\frac{w^3}{v^3}
	\frac{1}{K(a,b)},
	\label{eq:polaron_mobility}
\end{equation}
where $\beta=\hbar\omega_{\mathrm{LO}}/(k_{\mathrm{B}}T)$, and $v$ and $w$ are temperature-dependent variational parameters obtained by minimizing the polaron free energy. The function $K(a,b)$ accounts for the relevant variational and temperature-dependent terms in the Feynman--Hellwarth mobility expression; its explicit definition and the associated parameters are provided in the SM. The mobility was evaluated using the Gaussian-cgs form of the expression, with all quantities and conversion factors treated consistently in this unit system. The resulting mobilities are reported in $\mathrm{cm}^2\,\mathrm{V}^{-1}\,\mathrm{s}^{-1}$ in Table~\ref{tab:polaron}. The optimized Feynman variational parameters ($v$ and $w$) for electrons and holes are provided in Table~S10 of the SM.

The calculated electron-polaron mobilities range from 39.36 to 99.50~cm$^2$~V$^{-1}$~s$^{-1}$ at room temperature. Among the investigated structures, tetragonal Sr$_5$AsBi(NH)$_2$ exhibits the highest calculated mobility, whereas tetragonal Ca$_5$AsSb(NH)$_2$ has the lowest. The higher mobility of the former is consistent with its lower carrier effective mass and weaker Fr\"ohlich coupling. The corresponding hole-polaron properties, including their mobilities, are provided in Table~S9 of the SM; the electron-polaron parameters are summarized in Table~\ref{tab:polaron}. Previous work reported room-temperature carrier mobility estimates using a different transport treatment and a specified carrier concentration~\cite{chau2025defect}. The present calculations instead examine carrier transport within the Feynman--Hellwarth Fr\"ohlich-polaron framework, thereby providing a complementary assessment of the role of polar optical-phonon coupling. For both compositions with two investigated polymorphs, the calculated electron-polaron mobilities remain comparable between the tetragonal and monoclinic structures. This indicates that the structural distortions considered here have a relatively modest effect on the model-estimated Fr\"ohlich-polaron mobility, in contrast to their more pronounced influence on the exciton wavefunction distributions.

Overall, the calculations indicate weak-to-intermediate carrier--phonon coupling and moderate polaron mass enhancement within the Fr\"ohlich model used here. The comparison between Ca$_5$AsSb(NH)$_2$ and Ca$_5$AsBi(NH)$_2$ shows that chemical substitution can modify the Fr\"ohlich coupling and the resulting mobility. Taken together with the excitonic results, these findings highlight the complementary roles of lattice polarization in screening electron--hole interactions and renormalizing individual carrier properties. They also show that, within the models employed here, exciton wavefunction distributions are more sensitive to the considered structural distortions than the calculated electron-polaron mobilities.

\section{Conclusion}

In summary, our first-principles many-body investigation of the defect-imide double antiperovskites AE$_5$AsPn(NH)$_2$ (AE = Ca, Sr; Pn = Sb, Bi) provides a comparative picture of their electronic, excitonic, and carrier-transport properties across tetragonal $P4/mmm$ and monoclinic $Cm$ structural models. The electronic structures reflect a competition between spin--orbit coupling and quasiparticle self-energy corrections: SOC reduces the band gaps, particularly for the Bi-containing compounds, whereas the $G_0W_0$ corrections substantially increase them. The quasiparticle calculations also reveal changes in the orbital character of the band-edge states, highlighting the importance of treating relativistic effects and many-body corrections consistently.

The mBSE calculations yield exciton binding energies of 140--210 meV across the investigated structures. A central result is the partial decoupling between exciton energetics and wavefunction geometry. Chemical substitution modifies the binding energies and the corresponding hydrogenic estimates of exciton spatial extent. For compositions with both polymorphs, however, the structural distortion from $P4/mmm$ to $Cm$ produces comparatively modest changes in binding energy and estimated radius, while the real- and momentum-space exciton distributions show more pronounced modifications. The symmetry reduction from point group $D_{4h}$ to $C_s$ changes the symmetry constraints on electronic-state mixing and is associated with this reshaping of the excitonic distributions. Because the distortion also changes local coordination and electronic dispersion, the results establish a symmetry-associated structural effect rather than isolating symmetry as the sole causal variable. The main implication is that exciton binding energy alone does not fully characterize the sensitivity of an excitonic state to structural polymorphism.

Lattice polarization provides two complementary contributions to the excited-state and carrier properties. Its ionic dielectric response screens the electron--hole interaction, reducing the estimated exciton binding energies by approximately 9--16\% to 118--190 meV within the analytical phonon-screening model. At the same time, Fr\"ohlich coupling between individual carriers and LO phonons produces polaronic dressing, with calculated mass enhancements of 1.33--1.54 and spatially extended polaron radii. Within the Feynman--Hellwarth model, the estimated electron-polaron mobilities range from 39 to 99 cm$^2$ V$^{-1}$ s$^{-1}$. Although these mobilities vary with chemical composition, the differences between the tetragonal and monoclinic polymorphs are comparatively modest within this model.

Overall, the calculations identify a distinct hierarchy of structural sensitivity: the exciton wavefunction distributions respond more strongly to the considered polymorphic distortion than do the exciton binding energies or the model-estimated electron-polaron mobilities. This symmetry-associated reshaping of exciton geometry, alongside the complementary effects of lattice polarization on electron--hole screening and carrier renormalization, provides the central physical picture emerging from this study. More broadly, the results emphasize the value of examining exciton wavefunctions in addition to their energies when assessing the response of structurally complex semiconductors to polymorphism.

\begin{acknowledgments}
	D.M. acknowledges the Indian Institute of Technology Delhi for the award of a Junior Research Fellowship. S.B. acknowledges financial support from the Science and Engineering Research Board (SERB) through the Core Research Grant (Grant No. CRG/2019/000647), which supported the establishment of the High Performance Computing (HPC) facility ``Veena'' at IIT Delhi. The authors acknowledge the computational resources provided by the Veena HPC facility. We acknowledge ``Tejas'' IoE HPC facility of Dept. of Physics, IIT Delhi.
\end{acknowledgments}

\section{Data Availability Statement}
The data supporting the findings of this study are available within the article and its
Supplementary Material (SM). Additional data are available from the corresponding author
upon reasonable request.
\bibliographystyle{apsrev4-2}
\bibliography{ref}

\end{document}